\documentstyle[sprocl,epsfig]{article}

\def\Journal#1#2#3#4{{#1} {\bf #2}, #3 (#4)}

\def\NPA{{\em Nucl. Phys.} A}
\def\PLB{{\em Phys. Lett.}  B}
\def\PRL{\em Phys. Rev. Lett.}
\def\PRC{{\em Phys. Rev.} C}
\def\PRD{{\em Phys. Rev.} D}

\def\fot{\frac{1}{2}}

\def\ra{\rightarrow}

\def\al{\alpha}

\def\be{\begin{equation}}
\def\ee{\end{equation}}
\def\bea{\begin{eqnarray}}
\def\eea{\end{eqnarray}}

\begin{document}

\title{ON THE MUON CAPTURE IN $^{28}Si$ AND EXTRACTION OF $g_P$}

\author{S. CIECHANOWICZ}

\address{Institute of Theoretical Physics, Pl. Maksa Borna 9, 
50 204 Wroclaw,\\ Poland\\E-mail: ciechano@ift.uni.wroc.pl}

\author{F. C. KHANNA}

\address{Department of Physics, Theoretical Physics Institute, 
University of Alberta,\\
Edmonton, Alberta, Canada, T6G 2J1\\
and\\TRIUMF, 4004 Wesbrook Mall, Vancouver, B.C., Canada, V6T 2A3\\
E-mail: khanna@phys.ualberta.ca}

\author{E. TRUHL\'{I}K}

\address{Institute of Nuclear Physics, Czech Academy of Sciences,\\
250 68 \v{R}e\v{z} n. Prague,\\
Czech Republic\\E-mail: truhlik@ujf.cas.cz}
\maketitle\abstracts{ Comparison of the new data for the 
negative muon capture in $^{28}Si$ with the recent shell model calculations
using a full 1s-0d model space and the USD empirical effective 
interaction and the one-nucleon weak current provides the induced
pseudoscalar $g_P$ substantially smaller than  the value predicted
by PCAC and pion-pole dominance. We find that adding 
the effect from the one-pion exchange axial charge density and 
using the same scheme of calculations does not change  situation
to the value of $g_P$. }

\section{Introduction}

The four-current describing the axial part of the weak interaction 
of the muon and the proton consists of two parts \cite{BS}
\be
J^a_{A,\,\mu}(p',p,q) = i\bar{u}(p')[\,-g_A(q)\gamma_\mu \gamma_5\,
+\,i\frac{g_P(q)}{m_\mu}q_\mu\gamma_5\,]\,\frac{\tau^a}{2}\,u(p)\,.
\label{JA}
\ee
The constant of the induced pseudoscalar, $g_P$, is given by PCAC, 
pion-pole dominance and the Goldberger-Trieman relation as
\be
g_P(q) = \frac{2Mm_\mu}{m^2_\pi+q^2}g_A(q)\,, \label{gP}
\ee
where $M$, $m_\pi$ and $m_\mu$ is the nucleon, pion and muon mass 
respectively and $g_A(0)=-1.2601\pm 0.0025$ \cite{PDG}.

It is known that the experimental value of $g_P$ is the least known 
constant of the four constants ($g_V$,\,$g_A$,\,$g_W$,\,$g_P$) defining
the weak nucleon current. The best measurement of the ordinary muon
capture by the proton \cite{B1} yields $g_P$ with an uncertainty
of $42\%$ and the world average reduces this to $22\%$ \cite{B2}.
The recent precise measurement \cite{VOR} of the transition rate for the
muon capture by $^{3}He$  leads to the extraction \cite{CT}
of $g_P$ with an accuracy of $19\%$.

\section{Studies of $g_P$ in $A\,=\,28$ Nuclei}

\subsection{$\gamma-\nu$ Correlations}\label{subsec:gnc}

An interesting attempt is being made for many years, to study $g_P$
in the reaction
\be
\mu^- + ^{28}Si(0^+)\,\ra\,\nu_\mu + ^{28}Al^*(1^+;2201\,keV)\,
\ra\,\gamma + ^{28}Al^*(0^+;972.4\,keV)\,, \label{Reac}
\ee
by measuring the $\gamma-\nu$ correlation \cite{Ro,GHM,VB,Mo}.
The formula for the $\gamma-\nu$ correlations is \cite{CO,C96}
\be
W\,=\,1\,+\,a^0_2P_2(\hat{ k}\cdot\hat{ \nu})+
(\hat{ P}_\mu\cdot\hat{ k})(\hat{ k}\cdot\hat{ \nu})\,
[\,\al^0+\frac{2}{3}\,c^{\,0}_1+b^{\,0}_2\,]P_2(\hat{ k}\cdot\hat{ \nu})\,,
 \label{W}
\ee
and it is the coefficient $a^0_2$ which is measured. The correlation
coefficient $a^0_2$ is written in terms of reduced matrix elements
(r.m.e.) of multipole amplitudes between initial and final states as
\be
a^0_2\,=\,F\,\frac{1-x^2}{2+x^2}\,,\qquad x\,=\,\frac{|A_{1,\,fi}|}
{|M_{1,\,fi}|}\,,  \label{a02}
\ee
where the coefficient, $F=1$, is a function of the nuclear spin sequence 
in the $\gamma$-transition, $1^+\,\rightarrow\,0^+$.

The multipole operators $\hat{A}_1$ and $\hat{M}_1$ can be written 
in terms of the standard multipoles \cite{DW,Don84}
\bea
i\hat{A}_1\,& = & \,\frac{C}{\hat{J}_f}\,(\hat{\cal L}^A_1\,
-\,\hat{\cal M}^A_1)\,, \label{A1} \\
i\hat{M}_1\,& = & \,\frac{C}{\sqrt{2}\hat{J}_f}\,(\hat{\cal T}^{A,\,el}_1\,
-\,\hat{\cal T}^{V,\,mag}_1)\,.  \label{M1}
\eea
Since $a^0_2$ depends on the ratio of the r.m.e. it is expected that it
depends on the nuclear models only weakly. However, the calculations show
that this statement is only partially true.

The value of $a^0_2$ has recently been obtained \cite{Mo},
\be
a^0_2\,=\,0.360\,\pm\,0.059\,.  \label{a02exp}
\ee

\subsection{Impulse Approximation}\label{subsec:ia}

In order to compare the calculations with the result (\ref{a02exp}),
one should write down the currents. 
The leading terms of the one-nucleon weak vector and axial currents in 
the q-space are taken from \cite{CT}
\bea
\hat{\tilde {\vec{\jmath}}}_A(\vec q) & = & \tau_{-1} [\,g_A \vec \sigma\,-\,
\frac{g_P}{2M\,m_\mu}\, \vec q\,(\vec {\sigma} \cdot \vec q)\,]\,,
   \label{j1Aq}     \\
\hat{\tilde {\rho}}_A(\vec q) & = & \tau_{-1} [\,\frac{g_A}{2M}\,
\vec{\sigma}\cdot
(\vec{p}'+\vec{p})\,-\,\frac{g_P}{2M\,m_\mu}\,q_{\,0}\,(\vec{\sigma} \cdot 
\vec q)\,]\,, \label{rho1Aq} \\
\hat{\tilde {\vec{\jmath}}}_V(\vec q) & = & \tau_{-1} [\,\frac{g_V}{2M}
(\vec{p}'+\vec{p})\,+\,
\frac{g_V+g_M}{2M}\,i(\vec{\sigma} \times \vec{q})\,]\,. \label{j1Vq}  
\eea
Here $\vec{q}=\vec{p}'-\vec{p}$ and $q_0=m_\mu-\nu$.

The total weak current is equal to the sum of the
vector and axial currents.

In the configuration space,
\bea
\hat{\vec{\jmath}}_A(\vec x) & = & \tau_{-1} [\, g_A \vec{\sigma}\,
\delta(\vec{r} - \vec{x})\,+\,\frac{g_P}{2M\,m_\mu}\,\nabla_{\vec{x}}\,
(\nabla_{\vec{x}} \cdot \vec{\sigma}\,\delta(\vec{r}-\vec{x}))\,]\,,
\label{j1Ar} \\
\hat{\rho}_A(\vec x) & = & \tau_{-1} [ \frac{g_A}{2M} 
\{(\vec{\sigma} \cdot \vec{p})\,,
\delta(\vec{r} - \vec{x})\}-i\frac{g_P}{2M\,m_\mu}q_{\,0}
(\nabla_{\vec{x}} \cdot \vec{\sigma} \delta(\vec{r}-\vec{x}))]\,,
\label{rho1Ar} \\
\hat{\vec{\jmath}}_V(\vec x) & = & \tau_{-1} [\, \frac{g_V}{2M} \
\{\,\vec{p}\,,\,\delta(\vec{r} - \vec{x})\,\}\,+\,\frac{g_V+g_M}{2M}\,
(\nabla_{\vec{x}} \times \vec{\sigma}\,\delta(\vec{r}-\vec{x}))\,]\,.
\label{j1Vr}
\eea 
The Coulomb multipole, $\hat{\cal M}^A_{LM}$, for the one-nucleon weak axial charge 
density (\ref{rho1Ar}) and in the second quantized formalism 
\cite{Don84,W} is
\be
\hat{\cal{M}}^A_{LM}(1) \,= \,\int d^3x\, j_L(qx)Y_{LM}\,
\hat{\rho}_A(\vec x)
\,=\, \sum\limits_{\alpha'\,\alpha}\hat{M}^A_{LM}(\alpha',\alpha)\,
 c^{\dagger}_{\alpha'}\,c_{\alpha}\,.
\label{MALM}
\ee
The r.m.e. of the Coulomb multipole (\ref{MALM})
expressed in terms of the single particle r.m.e.'s and one-body
density-matrix elements as defined by Donnelly and Sick in Eq.\,(4.86) 
of Ref.\,\cite{Don84} is
\bea
M^A_{L,\,fi}(1) &\,\equiv \,& \big<f||\hat{\cal M}^A_{L}(1)||i\big>  = 
\nonumber  \\ 
& & \sum\limits_{j'j}\big<j';\fot||\hat{\cal M}^A_L(1)||j;\fot\big>
\big<f||{1\over\hat L\sqrt{3}}
\left[c^{\dagger}_{j'}\otimes{\tilde c}_j\right]_{L;1}||i\big>\,,
\label{rme1}
\eea
where 
\be
{\tilde c}_{jm\tau}\,=\,(-1)^{j+m}(-1)^{\fot+\tau}\,c_{j-m-\tau}\,.
\ee
Similar equations can be obtained for other multipoles entering 
Eq.\,(\ref{A1}) and Eq.\,(\ref{M1}).

\section{Exchange Charge Density}

Let us remind ourselves that
the spontaneous breaking of the chiral symmetry is accompanied by 
the appearance of pions as Goldstone bosons. The production and absorption
 of these bosons 
in the electroweak interactions on the nucleon is described 
by the low energy theorems \cite{AFFR,AD}. As a consequence, the time 
component of the one-nucleon and of the weak axial one-pion exchange 
currents are of the
same order in $1/M$. This fact makes the weak axial one-pion exchange
charge density a favourable object for studying the pionic degrees
of freedom in nuclei \cite{Tow}.

The leading term of the weak axial one-pion exchange charge density
is \cite{TK} 
\bea
\hat{{\tilde \rho}}_A(exch)\, & = & \,-i\sqrt{2}\,({\vec \tau}_1 \times 
{\vec \tau}_2)_{-1}\,
\big(\frac{g}{2M}\big)^2\,
\frac{m^2_\rho}{g_A}\,\Delta^\pi_F({\vec q}^{\,2}_2)\,
\Delta^\rho_F({\vec q}^{\,2}_1)\, \times\nonumber \\
& & F_{\pi NN}({\vec q}^{\,2}_2)\,F_{\rho NN}({\vec q}^{\,2}_1)\, 
({\vec \sigma}_2 
\cdot {\vec q}_2)\,+\,(1\,\longleftrightarrow\,2)\,, \label{MEC}
\eea
where
\be
\Delta^B_F({\vec q}^{\,2}_i)\,=\,\frac{1}{m^2_B\,+\,{\vec q}^{\,2}_i}\,,
\qquad\,
F_{B NN}({\vec q}^{\,2}_i)\,=\,\big(\frac{\Lambda^2_B\,+\,m^2_B}{\Lambda^2_B\,
+\,{\vec q}^{\,2}_i}\big)^{n_B}\,.
\ee
The parameters $\Lambda_B$ and $n_B$ from the
potential OBEPQB \cite{Mach} extended by the $a_1$ exchange
\cite{OPT} are used in the calculations.

The corresponding Coulomb multipole operator is
\bea
\hat{\cal{M}}^A_{LM}(exch) \,& = &\,\int d^3x\, j_L(qx)Y_{LM}\,
\hat{\rho}_A(\vec x;exch) \nonumber \\
\,& = & \, \sum\limits_{\alpha'_1\,\alpha'_2\,\alpha_1\,\alpha_2}
\hat{M}^A_{LM}(\alpha'_1,\alpha'_2,\alpha_1,\alpha_2;exch)\,
 c^{\dagger}_{\alpha'_1}\,c^{\dagger}_{\alpha'_2}\,c_{\alpha_2}\,
 c_{\alpha_1}\,.
\label{MALMEXCH}
\eea
The r.m.e. of the operator (\ref{MALMEXCH}) is defined again 
in accordance with Donnelly and Sick
(see Eq.\,(4.89) of Ref.\,\cite{Don84}),
\bea
\lefteqn{M^A_{L,\,fi}(exch)\,\equiv\,\big<i||{\hat {\cal M}}^A_L(exch)
||f\big> \null}\nonumber\\
&&\null = \sum\limits_{j'_1 j'_2 J' T' j_1 j_2 J T}
\big<\big[j'_1\otimes j'_2\big]_{\,J';\,T'}||
{\hat {\cal M}}^A_{L}(exch)||\big[j_1\otimes j_2\big]_{\,J;\,T}\big>\,\times
\nonumber\\
&&\null\big<i||{{-
1}\over\hat L\sqrt{3}}
\left\{\big[c^{\dagger}_{j'_1}\otimes c^{\dagger}_{j'_2}\big]_{J';T'}
\otimes\big[{\tilde c}_{j_1}\otimes{\tilde c}_{j_2}\big]_{J;T}
\right\}_{L;1}||f\big>\,.   \label{RMEEXCH}
\eea
As it is seen, the two-body matrix elements calculated in the 
single-particle basis are separated from the two-body density matrix 
elements. 

\section{Results and Discussion}

The results of the experiments \cite{Ro,GHM} 
have been compared to numerical results \cite{C76,PS} for specific 
nuclear models with contradictory conclusions. 
The comparison of the recent experimental data 
\cite{VB,Mo} for the correlation coefficient (\ref{a02}) with the 
results of the calculations \cite{C76} for the nuclear shell model 
\cite{Wil73,DeV73} has shown that these results yield the value of $g_P$ 
substantially lower in comparison with the PCAC prediction.  Moreover, 
the most recent analysis \cite{J} based on the wave functions of a full 
1s-0d model space obtained from the USD residual interaction with the 
OXBASH code \cite{OX} gives $g_P/g_A\,=\,0.0\pm3.2$.

Our calculations using the currents (\ref{j1Ar})-(\ref{j1Vr}) and the 
one-body density matrix elements obtained from OXBASH code with the W 
residual interaction are presented in Fig.\,1. Comparing the full 
curve with the full curve of Fig.\,3 of Ref.\,\cite{Mo} shows a weak
model dependence within one type of calculations. However, the 
results for various classes of models can differ considerably, yielding
quite a large model dependence of the extracted $g_P$. The dashed curve 
corresponds to $g_A(0)\,=\,-1$. This value of $g_A$ is advocated 
in Refs.\,\cite{MR,KTL,TOK,BP}. However, another approach
\cite{PAMG,KS} predicts only $5\%$-$10\%$ damping of $g_P$ in 
$A\,=\,28$ nuclei.
The dotted curve is obtained with $g_A(0)\,=\,-1$ and without the 
velocity dependent term in the one-nucleon axial charge density
(\ref{rho1Ar}). It is clear that the damping of the effect of this
term shifts the value of $g_P$ to the PCAC prediction.

\begin{figure}[t]
\center{\epsfig{file=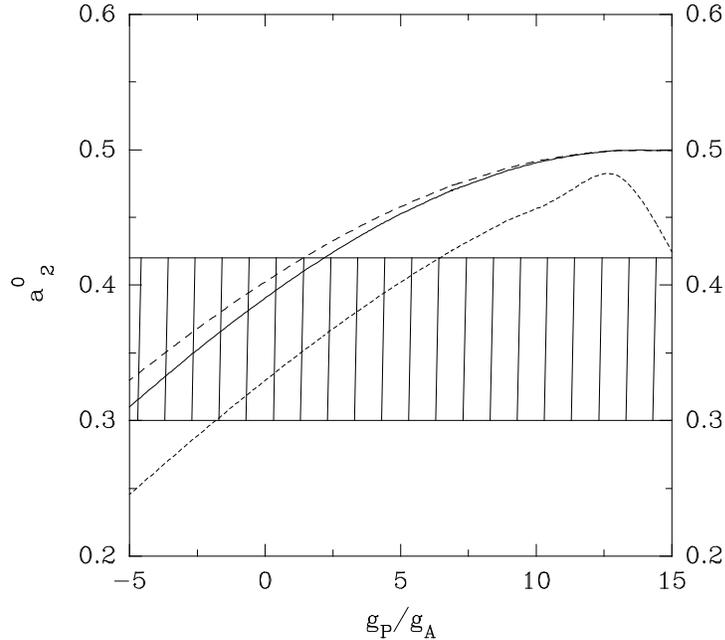,width=9.50cm}}
\caption{Dependence of the correlation coefficient $a^0_2$ on 
$g_P/g_A$; full curve corresponds to $g_A\,=\,-1.26$, dashed curve
is for $g_A\,=\,-1$ and the dotted curve is obtained with $g_A\,=\,-1$
and with the velocity dependent term in the one-nucleon axial charge
density omitted .  \label{fig:musi2}}

\end{figure}

In searching for a possible compensation of this velocity dependent term
we turned our attention 
to the fact that in the same type of the transition, $\Delta\,T\,=\,1$,
$0^+\,\ra\,1^+$, in $A=12$ nuclei \cite{GS}, the almost exact chiral 
symmetry of the strong interactions manifests itself via a large effect 
of the weak axial one-pion exchange charge density. Indeed, as it is 
seen from Table 6 of Ref.\,\cite{GS}, the matrix element of the weak axial 
charge density is enhanced after including the contribution of the soft 
pion exchange charge density by almost $40\%$, which is just an amount
demanded by the data. There is no reason
to believe that the effect of the exchange charge density does not take 
place also in the case
of reaction (\ref{Reac}), unless some other
large nuclear physics effect (core polarization etc.) does not come into
play and acts in a opposite direction.
The two-body density matrix elements were calculated again using  
the OXBASH code and the W residual interaction. However, in contrast to 
the case of $A=12$ nuclei \cite{GS},
in our scheme of calculations the effect of the weak axial one-pion 
exchange charge density (\ref{MEC}) turns out to be negligibly small.
This result may be attributed to the fact that the wavefunctions and 
the operators
of nuclear currents are not constructed consistently. This point seems to
be proved at least for the one-nucleon currents in the recent report
\cite{SSH}, where Siiskonen {\it et al} constructed consistently
renormalized one-particle transition operators and an effective
interaction starting from a realistic NN interaction
and the G-matrix appropriate for the 1s-0d shell. This leads to a
result in the range $4.4\,\le\,g_P / g_A\,\le\,5.9$ thus leaving 
space for an 
effect of $\approx\,30\%$ of the weak axial one-pion exchange 
charge density. Let us note that the authors of Ref.\,\cite{SSH}
use the value $g_A(0)\,=\,-1$.  

In conclusion we note that the present calculation fails to give 
a value of $g_P$ consistent with that obtained from PCAC and the pion-pole
dominance. A possible reason for this result may be traced to an 
inconsistent treatment of the wavefunctions and the transition operators
and to the fact that many of the usual renormalization effects have 
not been taken into account.

\section*{Acknowledgments}
This work was supported in part by the grant GA \v{C}R 202/97/0447.
The research of F.C.K. is supported in part by NSERCC. A part of the work
was done during the stay of E.\,T.\, at the University of Edmonton. He thanks
both TRIUMF and the Theoretical Physics Institute for their support
and Prof.\, F.\,C.\, Khanna for a warm hospitality.
 
We thank Prof. B. A. Brown for providing us with the one- and two-body
density matrices and Dr. R. K. Teshima for computing some radial 
integrals. The discussion with Prof. A. Thomas is acknowledged.

\end{document}